\documentclass[%
a4paper,
prd,
twocolumn,
superscriptaddress,
preprintnumbers,
nofootinbib,
nobibnotes,
 amsmath,amssymb,
 aps,
]{revtex4-2}
\usepackage{upgreek,amsmath,amsfonts,graphics,epsfig,amssymb,color}
\usepackage{graphicx}
\usepackage{dcolumn}
\usepackage{bm}

\newcommand{\black}{\color{black}}
\newcommand{\red}{\black}
\usepackage{hyperref}
\begin{document}
\preprint{INR-TH-2022-001}
\title{Muon lateral distribution function of extensive air showers: results of the Sydney University Giant Air-shower Recorder versus modern Monte-Carlo simulations}
\author{N.\,N.\,Kalmykov}
\affiliation{D.V.~Skobeltsyn Institute of Nuclear Physics,
M.V.~Lomonosov Moscow State University, Moscow 119991, Russia}
\author{I.\,S.\,Karpikov}
\email{karpikov@inr.ru (corresponding author)}
\author{G.\,I.\,Rubtsov}
\author{S.\,V.\,Troitsky}
\affiliation{Institute for Nuclear
Research of the Russian Academy of Sciences, 60th October Anniversary
Prospect 7a, Moscow 117312, Russia}
\begin{abstract}
The Sydney University Giant Air-shower Recorder (SUGAR) measured the muon component
of extensive air showers with a unique array of muon detectors.  The SUGAR data allows us to reconstruct
the empirical dependence of muon density on the distance from the axis of the shower, the lateral
distribution function (LDF). We compare the shape of this function with the predictions of hadronic-interaction
models, QGSJET-II-04 and EPOS-LHC, \textcolor{black}{in the energy range $10^{17.6}$ - $10^{18.6}$~eV}. We find a difference
between the observed data and the simulation: the observed muon density falls faster with the increased core distance than it is predicted in simulations. This observation may be important for interpretation of the energy-dependent discrepancies in the simulated and observed numbers of muons in air showers, known as the ``muon excess''.
\end{abstract}
\maketitle


\section{Introduction}
\label{sec:intro}
Ultra-high-energy cosmic rays (UHECRs) enable particle-physics studies beyond the capabilities of terrestrial colliders. However, due to their low flux, UHECRs can only be observed indirectly, via extensive air showers (EAS).
Hadronic interactions play an important role in the EAS development. Modeling of the hadronic component of EAS is important for studying the primary composition of cosmic rays \cite{Kampert_and_Unger}. The number of muons in EAS is directly related to the hadronic interactions. Recently, much attention has been paid to the discrepancies between the number of muons in theoretical models of the development of EAS, implemented in Monte-Carlo simulations, and in real EAS data, see e.g.\ Refs.\ \cite{Hans_factor,Dembinski:2021szp1} for reviews. Some experiments reported this ``muon excess'' at various primary energies, muon energies,zenith angles, atmospheric depths and core distances (SUGAR \cite{SUGAR-spark1, SUGAR-spark2, SUGAR_INR}, HiRes-MIA \cite{HiRes-MIA}, NEVOD-DECOR \cite{NEVOD-DECOR-2010,NEVOD-DECOR}, Yakutsk \cite{old-Yakutsk}, Pierre Auger Observatory and AMIGA \cite{Auger-mu, AMIGA-mu}, Telescope Array \cite{TA-mu}). However, other measurements, performed under different conditions, show the agreement in the muon number between data and models (EAS-MSU \cite{EAS-MSU_mu}, Yakutsk \cite{new-Yakutsk}, KASCADE-Grande \cite{KASCADE-Grande}, IceTop \cite{new-IceTop}).

Ref.\ \cite{Hans_factor} presented a joint analysis of the EAS muon content measured by various experiments (EAS-MSU, IceTop, KASCADE-Grande, NEVOD-DECOR, Pierre Auger Observatory, AMIGA, SUGAR, Telescope Array, Yakutsk). It has been shown that, on average, the difference between the observed and predicted muon densities grows with the primary energy. However, higher-energy events are rare and are recorded by larger installations, hence often the muon content of higher-energy EAS is measured at larger distances from the core than for
lower energies. Therefore, change of the shape of the muon \textcolor{black}{lateral distribution function (LDF)} and the energy dependence of its normalization may become degenerate. To study the muon LDF, an installation with a large array of muon detectors is best suited.

\textcolor{black}{In this work, we use the data of the Sydney University Giant Air-shower Recorder} (SUGAR, see e.g.\ Refs.~\cite{SUGAR1, Brownlee, SUGAR3}). The main SUGAR data set is based on observation of muons with shielded detectors only. However, surface-located spark chambers were used to study the electromagnetic component as well, and their results were probably the first demonstration of the muon excess in air showers \cite{SUGAR-spark1,SUGAR-spark2}. In a previous work \cite{SUGAR_INR}, we compared the cosmic-ray spectra measured by SUGAR~\cite{SUGARWinn} and by the Pierre Auger Observatory~\cite{PAO1, Aab:2017njo} and determined the empirical dependence of the number of muons in a vertical shower on the primary energy. This empirical relationship between primary energy and the number of muons was compared with that predicted from the hadronic models and it was found that the models predicted fewer muons. The purpose of the present work is to determine the experimental muon LDF using the SUGAR data and to compare it with the predictions of Monte-Carlo simulations.

The rest of the paper is organized as follows. In Sec.\  \ref{sec:SUGAR_array}, we discuss briefly the SUGAR array and the data used, including criteria of event selection. Monte-Carlo simulations are described in Sec.~\ref{sec:MC}. Section~\ref{sec:anal-res} describes the procedure of the LDF comparison between data and Monte-Carlo simulations and presents our results. We \red discuss systematic uncertainties of the method in Sec.~\ref{sec:syst} and \black briefly conclude in Sec.~\ref{sec:concl}. \textcolor{black}{Appendices \ref{sec:LDF} and \ref{sec:en_formuls} summarize technical information} from previous publications, while Appendix \ref{sec:Appendix} presents an update of the results of Ref.~\cite{SUGAR_INR} with a more detailed Monte-Carlo simulation performed for the present work and with a revised zenith angle range. 

\section{The SUGAR array and its data}
\label{sec:SUGAR_array}
The SUGAR experiment was in operation between 1968 and 1979 ~\cite{SUGAR1, Brownlee, SUGAR3} in New South Wales, Australia, at the altitude of $\sim 250$~m above sea level. The array covered an area of about 70 km$^2$ and consisted of 54 autonomous stations, each consisting of two underground detectors spaced 50 m apart in the North-South direction. Each detector was an underground tank containing a pool of liquid scintillator with an effective area of  6.0 m$^2$. The detector tanks were buried at the depth of 1.5 m \cite{Brownlee}.
These detectors, therefore, were intended to record only muons with energies above the threshold of
$0.75\sec\theta_{\mu}$ GeV, where $\theta_{\mu}$ is the zenith
angle of the incident muon.
However, each detector station had a maintenance hole on top of it, through which the muon signal was contaminated by the electromagnetic component of a shower for small zenith angles. It has been shown that this contamination is negligible for showers with zenith angles $\theta >17^\circ$ \cite{LSW-thesis}, and in what follows, we include only such events in the analysis. The minimum measured  density in the detector was 2.4 vertical equivalent muons per its area of 6~m$^2$. Each station was triggered and recorded a "local event" when the density in both detectors exceeded 2.4 vertical equivalent muons. The records from the stations were compared with records from all other stations and the presence of an air shower event was registered when three or more stations could be
triggered by the passage of an air shower front through the array. 
There were 13729 such events during the eleven years of operation of the array


For this work, we use the detailed data on these events which include readings of individual detector stations. The following selection criteria, similar to the standard SUGAR analysis \textcolor{red}{ \cite{SUGAR1, Brownlee, SUGAR3} }, were imposed:
\begin{enumerate}
    \item
 The shower axis is located in a square constrained with $|X|$ and $|Y| < 5000$ m.
 \item
 Events are removed in which the triggered detector is located at a distance from the shower axis of more than 5000~m.
 \item
 Events with saturated detector stations, $>4000$ particles, are removed.
 \item
 Reconstructed zenith angles are in the range $17^{\circ}<\theta\leq70^{\circ}$.
 \end{enumerate}
 For each of the events, we determine the effective number of ``vertical muons'', $N_v$, using the standard SUGAR procedure, see Appendix~\ref{sec:LDF}. Then we make use of the result of Ref.~\cite{SUGAR_INR}, which determined a relation between $N_v$ and the reconstructed primary energy, $E$, from the normalization to the Auger spectrum (see Appendix~\ref{sec:en_formuls} for \red details\black). For the present study, we select events with $10^{17.6}<E<10^{18.5}$ eV (this is the only place where the determination of $E$ is used). The lower energy limit is due to the fact that at  $E<10^{17.6}$ eV, the event registration efficiency becomes less than 50\%. The upper energy limit is
 determined by the low statistics of high-energy events. With these selection criteria, $N_{\rm data}=4514$ events remain, which we use to analyze muon LDF.

\section{Monte Carlo Simulation.}
\label{sec:MC}
We  use  the CORSIKA 7.4001 \cite{Heck:1998vt}
EAS
simulation package. We choose the QGSJET-II-04~\cite{Ostapchenko:2010vb} or
EPOS-LHC~\cite{EPOS-LHC} as the high-energy
hadronic interaction
models, and FLUKA2011.2c~\cite{Fluka} as the low-energy hadronic
interaction model.
For each of the two high-energy hadronic interaction models, we simulated 15000 showers for primary protons and the same number of showers for primary iron, with thrown primary energies following an $E^{-3.19}$ differential spectrum
\cite{SUGARWinn} with
$9\times10^{16}$~eV$<E<4\times10^{18}$~eV and with zenith angles in the range between $17^{\circ}$ and $70^{\circ}$,
assuming an isotropic distribution of arrival directions in the celestial sphere.
The simulations were performed with the thinning parameter
$\epsilon= 10^{-5}$ with the limitation of the maximum weights according to Ref.~\cite{Kobal}.

While Monte-Carlo models of modern experiments include detailed simulations of the detector response to the EAS particles, we find this inappropriate for the present analysis because the available information about the detector and electronics is insufficient for construction of a reliable response model. Instead, we use a traditional approach, where we calculate the expected number of muons at each detector of a station and determine the triggered status of that station by assuming Poisson fluctuations.

\begin{figure}
\centerline{
\includegraphics[width=0.95\linewidth]{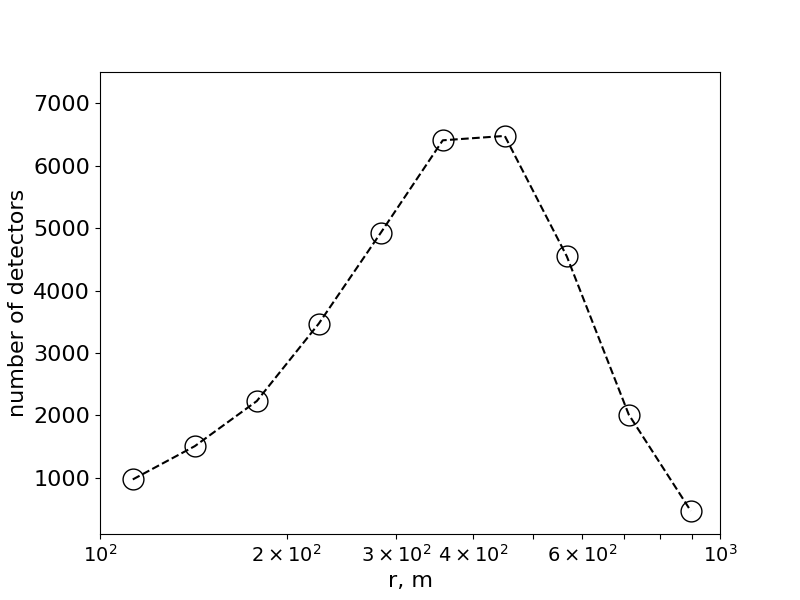}}
\caption{Distribution of triggered detector stations versus distance from the shower axis. \red Each point gives the total number of triggered stations of the 4514 selected events in the core-distance bin.}
\label{fig:det_dis}
\end{figure}

For each simulated EAS, we select muons with energies above the detector threshold and calculate their mean numbers in concentric rings around the shower axis, the range 100~m$\le r \le 1000$~m from the axis in 10 bins with widths equal in terms of $\log (r)$. These mean densities are used as Poisson means to produce 10 readings at artificial ``stations'', one in each ring. \textcolor{black}{Stations} with readings below the threshold of 2.4 vertical equivalent muons were discarded. In addition we have discarded certain number of \textcolor{black}{stations} to reproduce the distribution of detectors in $r$ observed in real data, see Fig.~\ref{fig:det_dis}.
The ensemble of these artificial events was treated in the same way as the real data were treated. 

\section{Analysis and results}
\label{sec:anal-res}
In this work, we concentrate on the LDF shape and not on its normalization. Hence, for each event, either real or artificial, we normalize all measured muon densities to the effective total number of muons, $N_\mu$, determined with the standard SUGAR procedure, see Appendix~\ref{sec:LDF} (note that this normalization is the only step in the analysis where we use the LDF expression adopted by SUGAR). The muon density is multiplied by the normalization factor $10^{6.6}/N_\mu$. We then consider ensembles of real and simulated normalized individual station readings and no longer use the information about EAS events to which these readings were associated. The normalized readings are binned in the core distances in the same 10 bins we introduced for the Monte-Carlo showers. Then we compare these normalized binned LDF in the data with the MC. The accuracy in determining  $N_\mu$ in Monte Carlo is $\log_{10}(\Delta N_\mu$) = 0.18.

\begin{figure}
\begin{minipage}[h]{0.95\linewidth}
\center{\includegraphics[width=1.\linewidth]{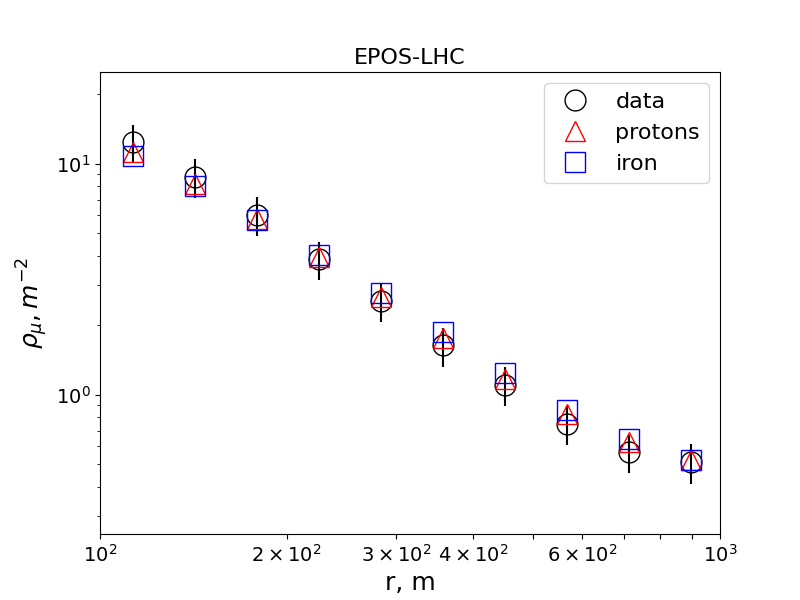} \\ a)}
\end{minipage}
\hfill
\begin{minipage}[h]{0.95\linewidth}
\center{\includegraphics[width=1.\linewidth]{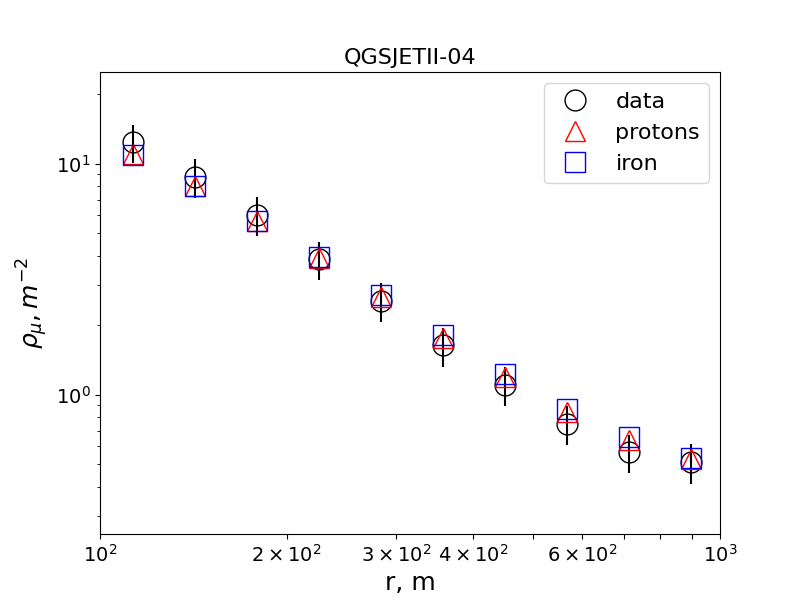} \\
b)}
\end{minipage}
\caption{Mean muon lateral distribution function (LDF). Black open circles with error bars
indicate the muon LDF of data; red open triangles indicate muon LDF of proton Monte-Carlo; blue open squares  indicate muon LDF of iron Monte-Carlo. Figure a) shows the  Monte-Carlo simulations with EPOS-LHC; figure b) Monte-Carlo simulations with QGSJET-II-04.}
\label{fig:LDF}
\end{figure}

Fig.~\ref{fig:LDF}
presents this comparison. We see that the overall agreement between data and simulations is reasonable. Note that at large $r$, the quantities compared and presented in Fig.~\ref{fig:LDF} do not represent the true LDF because we did not account for sub-threshold detector stations: the data do not contain information on whether the particular station is operation at the EAS arrival moment. Zero detector readings were consistently ignored both in data and in MC. This explains the behavior of the function at large $r$.

To estimate quantitatively the agreement between the data and simulations, we calculate the $\chi^{2}$ value,
\begin{equation}
\chi^{2}=\sum\limits_{i=1}^{\mathrm{ \textcolor{black}{bins} }}\frac{N_{\rm data}(P_{i}^{\rm data}-P_{i}^{\rm mc})^{2}}{P_{i}^{\rm data}},
\label{chi2}
\end{equation}
where $P_{i}^{\rm data}$  and $P_{i}^{\rm mc}$ are the bin value of the LDF function normalized to one, cf.\ Fig.~\ref{fig:LDF}, $N_{\rm data}=4514$  is the total number of events involved in the analysis. For EPOS-LHC $\chi^{2}/\mbox{d.o.f.}$ = 1.9 and 3.8 for proton and iron, respectively; for QGSJET-II-04 $\chi^{2}/\mbox{d.o.f.}$ = 2.3 and 3.5 for proton and iron, respectively.

\begin{figure}
\begin{minipage}[h!]{0.95\linewidth}
\center{\includegraphics[width=1\linewidth]{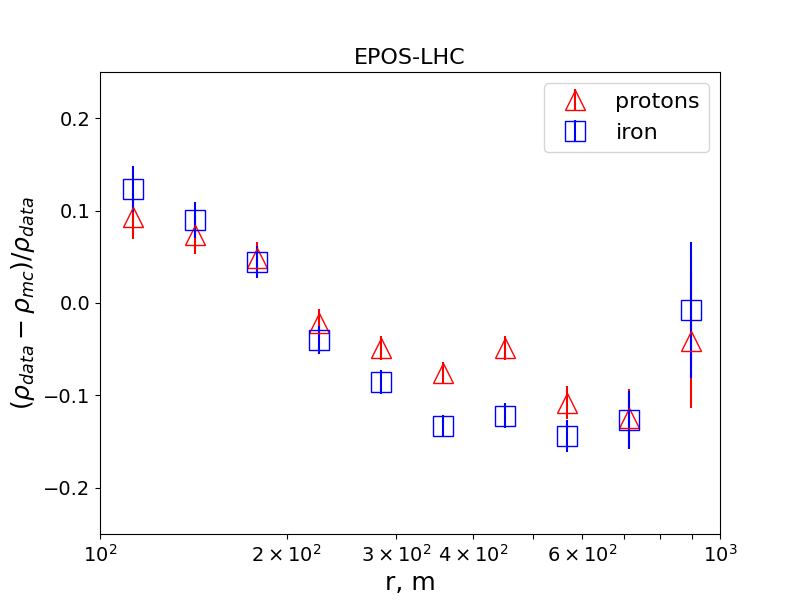}
\\
a)}
\end{minipage}
\hfill
\begin{minipage}[h!]{0.95\linewidth}
\center{\includegraphics[width=1\linewidth]{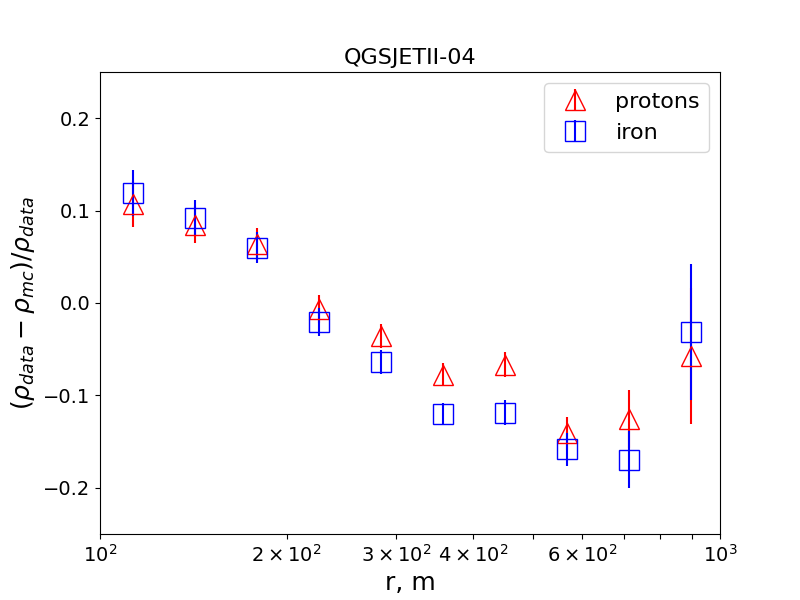} \\
b)}
\end{minipage}
\caption{Fractional difference between the experimental and simulated muon LDF. Red open triangles indicate muon LDF of proton Monte-Carlo; blue open squares  indicate muon LDF of iron Monte-Carlo. Error bars correspond to statistical uncertainly. Panel a): the  Monte-Carlo simulations with EPOS-LHC; panel b): with QGSJET-II-04.}
\label{fig:delta_LDF}
\end{figure}

We observe some difference between the experimental data and Monte Carlo: at close distances, the muon density in the data is higher than in Monte-Carlo; at large distances, on the contrary, the muon density in the data is less than in Monte-Carlo. The ratio of the observed and simulated LDF demonstrates this clear trend, better seen in Fig.~\ref{fig:delta_LDF}.
The simulations underestimate the LDF slope: more muons are concentrated at $r\lesssim 200$~m in the data than is predicted by models. This effect is statistically significant, with the probability to be caused by chance $p \approx 0.007$, estimated by the Pearson's chi-squared method. This trend is present for both hadronic-interaction models and both primary particle types we considered.

\red
\section{Systematic uncertainties}
\label{sec:syst}
In the present study, we concentrate only on the shape of the LDF, which reduces considerably systematic uncertainties, compared to studies of the LDF normalization, the ``muon excess''. Let us discuss briefly potential sources of systematics.

\paragraph{Energy scale.}
We discuss the energy determination and corresponding uncertainties in Ref.~\cite{SUGAR_INR} and summarize the discussion in Appendix~\ref{sec:en_formuls}. As we have already pointed out, for the present analysis, the primary energy is used only to select the events for the data set. The uncertainty of the energy scale translates into the uncertainty of the range of energies to which our results are applicable, but does not affect the results themselves since we are studying the shape of the normalized LDF. The change of the energy scale might manifest itself only if the shape of the muon LDF depended strongly on the primary energy within the energy range of interest. We tested that this is not so by repeating our analysis for two parts of the data set, with primary energies $10^{17.6}~\mbox{eV}\le E < 10^{18.0}~\mbox{eV}$ and $10^{18.0}~\mbox{eV}\le E \le 10^{18.5}~\mbox{eV}$. The difference in the LDF shape between the two subsets is negligible and does not exceed 0.01 in terms of the fractional difference plotted in Fig.~\ref{fig:delta_LDF}.

\paragraph{Detector calibration.}
Possible systematic errors in the measurements of the muon density may be related to the absolute detector calibration. However, in the present study we operate with the normalized
LDF; since the errors of this class affect the overall LDF scale, they cancel at normalization.

\paragraph{Nonlinearity.}
However, in case there exist certain nonlinearity related to the saturation effects, it may affect the LDF shape because the saturated detectors are always located at small core distances. We cannot find a detailed description of the nonlinearity in SUGAR detectors and therefore in our analysis we excluded all events for which at least one detector was saturated, see Sec.~\ref{sec:SUGAR_array}. To test the potential effect of nonlinearities, we lowered the saturation threshold by a factor of two and repeated our analysis. The change in the results was negligible compared to the statistical errors.
\black

\section{Conclusions}
\label{sec:concl}
We have used the data on muon content of EAS caused by primary cosmic-ray particles with energies $\sim 10^{18}$~eV and recorded by SUGAR to study the shape of the muon LDF and to compare it with the predictions of hadronic-interaction models EPOS-LHC and QGSJET-II-04.
\red
We found that both models predict a slower decrease of the muon density with increasing core distance for both proton and iron primaries
\black
than it is observed in the data, see Fig.~\ref{fig:delta_LDF}. We also used improved Monte-Carlo simulations performed for this work to update our previous results on the muon content of SUGAR-detected EAS reported in Ref.~\cite{SUGAR_INR}, see Appendix~\ref{sec:Appendix}.

\appendix
\section{Normalization}
\label{sec:LDF}
The muon number, $N_{\mu}$, is determined by fitting individual detector readings
by the experimentally determined muon lateral distribution function (LDF) \textcolor{black}{\cite{SUGAR-spark1, SUGAR-spark2} },
\begin{equation}
\rho_\mu(r) = N_{\mu}\,k(\theta)\, \left(\frac{r}{r_0}\right)^{-a}\,\left(
1 + \frac{r}{r_0}\right)^{-b}\,\,.
\label{Eq:LDF}
\end{equation}
Here, $\rho_{\mu }$ is the muon density, $N_{\mu}$ is the estimated total
number of muons, $\theta$ is the incident zenith
angle, $r$ is the perpendicular distance from the shower axis, $r_0 =
320$~m, $a = 0.75$, $b= 1.50 + 1.86 \cos\theta$, and
\begin{equation}
k(\theta) = \frac{1}{2\pi r_0^2}\,
\frac{\Gamma(b)}{\Gamma(2-a)\,\Gamma(a+b-2)}\,.
\end{equation}


\section{Energy estimation}
\label{sec:en_formuls}
 For a given EAS zenith angle $\theta$, the
effective vertical muon number $N_{\rm v}$ in a shower is related to
the
reconstructed muon number $N_{\mu }$ through the following relation \cite{SUGAR_Nv},
\begin{equation}
\begin{array}{l}
\log_{10}\left(\frac{N_{\rm v}}{N_{\rm r}}\right)=
(1-\gamma_{\rm v})^{-1} \cdot \\
\left(1-\gamma_{\rm v}-A(\cos\theta-1))
\log_{10}\left(\frac{N_{\mu}}{N_{\rm r}}\right)+\right. \\
\left.
B(\cos{\theta}-1)+\log_{10}
\left(\frac{1-\gamma_{\rm v}}{1-\gamma_{\rm v}-A(\cos\theta-1)}\right)
\right)
,
\end{array}
\label{Eq:Nv_Nmu}
\end{equation}
where the coefficients are $A=0.47$, $B=2.33$,
$\gamma_{\rm v}=3.35$, and the normalization scale is $N_{\rm r}=3.16\times10^{7}$.

The primary energy of a shower is related to $N_{\rm v}$
by the following expression,
\begin{equation}
E=E_{\rm r}(N_{\rm v}/N_{\rm r1})^{\alpha},
\label{Eq:E}
\end{equation}
where $N_{\rm r1}=10^{7}$, parameters $\alpha=1.018\pm 0.0042_{\rm stat}\pm 0.0043_{\rm
syst~SUGAR}\pm0.0028_{\rm syst~Auger}$ and $E_{\rm r}=(8.67 \pm 0.21_{\rm stat} \pm 0.26_{\rm syst~SUGAR}\pm1.21_{\rm syst~Auger})\times10^{17}~\mbox{eV}$ are obtained in Ref.\ \cite{SUGAR_INR} from the comparison of the SUGAR and Auger spectra.
\red
Parameters $\alpha$ and $E_{\rm r}$ were determined \cite{SUGAR_INR} from the requirement that the SUGAR spectrum matches the spectrum observed by Auger, which is justified since both experiments have similar fields of view in the Southern hemisphere.
As it is discussed in Ref.~\cite{SUGAR_INR},
the main systematic uncertainty comes therefore from the systematic error of 14\% in the Auger energy scale \cite{PAO1}.
\black

\begin{figure}
\begin{minipage}[h]{0.95\linewidth}
\center{\includegraphics[width=1.\linewidth]{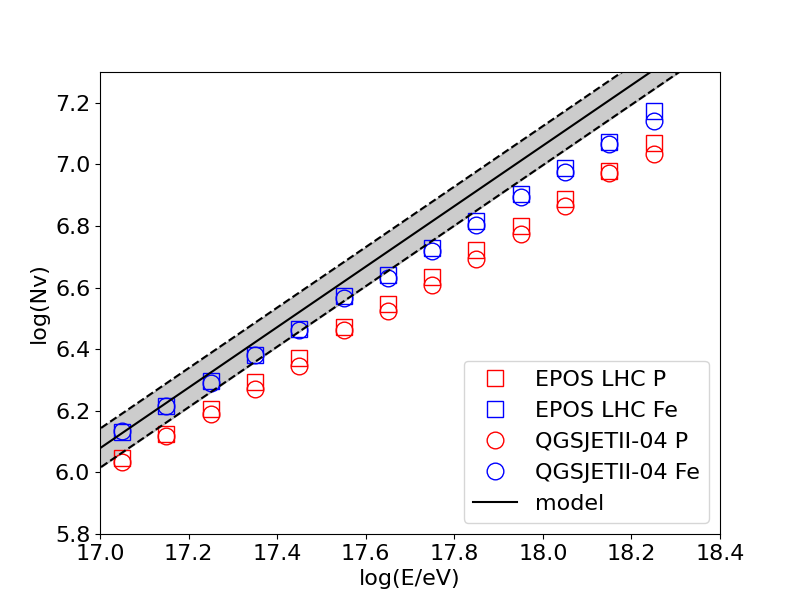} \\ a)}
\end{minipage}
\begin{minipage}[h]{0.95\linewidth}
\center{\includegraphics[width=1.\linewidth]{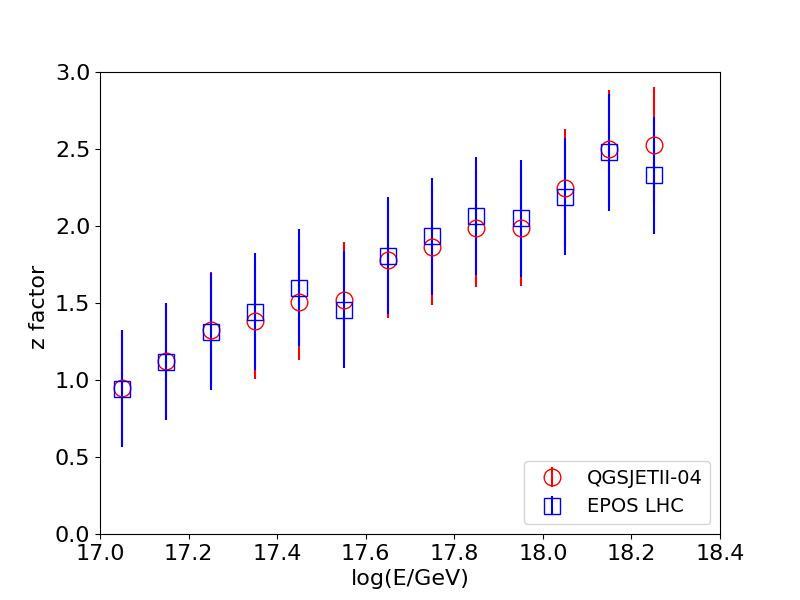} \\ b)}
\end{minipage}
\caption{Update of the results of Ref.~\cite{SUGAR_INR}. a) Mean effective number of vertical muons $N_{\rm v}$ as a function of the primary energy. Points
indicate the results of Monte-Carlo simulations with QGSJET-II-04 (protons - red open circles,
iron - blue open circles), EPOS-LHC (protons - red open quadrature, iron - blue open quadrature). The  gray line corresponds to our empirical model (\ref{Eq:E}); the shaded gray area indicates the total uncertainly (statistical and systematic errors\red, see the text, \black summed in quadrature). 
b) The $z$ factor versus the primary energy. }
\label{fig:Nv_vs_E}
\end{figure}

\section{Update on the results of Ref.~\cite{SUGAR_INR}}
\label{sec:Appendix}

Monte-Carlo simulations used in Ref.~\cite{SUGAR_INR} have been improved and corrected for the present work. The new simulations differ in the account of the threshold of the detector signal and the fluctuation of the muon density. In addition, Ref.~\cite{SUGAR_INR} used all zenith angles $\theta \le 70^\circ$, which might introduce a bias for near-vertical showers because of the detector maintenance hole discussed above. Here, we present an update of Ref.~\cite{SUGAR_INR} with the new simulations and for $17^\circ \le \theta \le 70^\circ$. In brief, we use the experimental muon LDF (\ref{Eq:LDF}) and fit it to the distribution of the muon density in the MC, obtaining $N_{\mu}$. Then we use Eq.~(\ref{Eq:Nv_Nmu}) to  express the effective number of vertical muons $N_{\rm v}$ in terms of $N_{\mu}$ and $\theta$. As a result, $N_{\rm v}$ is determined for each  shower \textcolor{black}{(both in data and in simulations)}. Thus we can plot the dependence of the number of vertical muons on the primary energy and compare it with Eq.~(\ref{Eq:E}) obtained \cite{SUGAR_INR} from comparison of the SUGAR and Auger spectra.

To compare the experimental muon density with the simulation predictions, we use the $z$ factor \cite{Hans_factor},
\begin{equation}
z = \frac{\log_{10}(N_{\rm v}^{\rm SUGAR})-\log_{10}(N_{\rm v}^{\rm p})}{\log_{10}(N_{\rm v}^{\rm Fe})-\log_{10}(N_{\rm v}^{\rm p})},
\end{equation}
where $N_{\rm v}^{\rm SUGAR}$ is the number of vertical muons calculated by the formula (\ref{Eq:E}), $N_{\rm v}^{\rm p}$ and $N_{\rm v}^{\rm Fe}$ are the simulated mean numbers of vertical muons for proton and iron primaries.

Fig. \ref{fig:Nv_vs_E}(a) presents a comparison of the new simulation results for $N_{\rm v}(E)$ with our empirical model (\ref{Eq:E}). Fig.\
\ref{fig:Nv_vs_E}(b) shows the dependence of the $z$ factor on energy. Compared with our previous work, the $z$ factor is smaller, so that the disagreement in the total muon number between the data and simulations, ``the muon excess'', is more modest. \red The most probable physical reason for this difference is the contamination of the muon signal by the electromagnetic component for vertical showers, which we did not take into account in Ref.~\cite{SUGAR_INR}. Since SUGAR measured only muons, potential shifts of the energy scale are degenerate with the shifts in muon number; this systematic uncertainty is dominated by the uncertainty of the Auger energy scale, see Appendix~\ref{sec:en_formuls}. Note that this uncertainty does not affect the results of the present work which studies the shape, and not the normalization, of the LDF, see Sec.~\ref{sec:syst}.
\black

\acknowledgements
We are indebted to Juris Ulrichs for his help and support crucial for our work with the SUGAR data and for intense discussions of the analysis. The results of this paper have been presented and discussed in the Working Group on Hadronic Interactions and Shower Physics (WHISP), to participants of which we are indebted for interesting comments.
We thank Leonid Bezrukov for helpful discussions \red and the anonymous reviewer for careful reading of the manuscript and numerous helpful comments\black. Monte-Carlo simulations
have been performed at the computer cluster of the Theoretical Physics Department, Institute for Nuclear Research of the Russian Academy of Sciences. Development of the analysis
methods (IK and GR) was supported by the Russian Science Foundation (grant 17-72-20291).

\bibliography{SUGAR}

\end{document}